\documentclass[preprint,5p,superscriptaddress]{elsarticle}

\usepackage{lineno,hyperref}
\usepackage{graphicx} 
\usepackage{textcomp}
\modulolinenumbers[5]

\journal{Journal of Alloys and Compounds}

\biboptions{square,comma,numbers,sort&compress}
\begin{document}

\begin{frontmatter}

\title{On the shear-affected zone of shear bands in bulk metallic glasses}

\author[IMP]{Farnaz A. Davani}
\author[IMP]{Sven Hilke}
\author[IMP]{Harald R\"osner\corref{correspondingauthor}}
\ead{rosner@uni-muenster.de}
\author[IFW]{David Geissler}
\author[IFW]{Annett Gebert}
\author[IMP]{Gerhard Wilde}

\cortext[correspondingauthor]{Corresponding author}

\address[IMP]{Institut f\"ur Materialphysik, Westf\"alische Wilhelms-Universität M\"unster, Wilhelm-Klemm-Str. 10, 48149 M\"unster, Germany}
\address[IFW]{Leibniz IFW Dresden, Institute for Complex Materials (IKM), Helmholtzstr. 20, 01069 Dresden, Germany}



\begin{abstract}
Notched bars of bulk metallic glasses, Pd$_{40}$Ni$_{40}$P$_{20}$ and Zr$_{52.5}$Cu$_{17.9}$Ni$_{14.6}$Al$_{10}$Ti$_{5}$, were deformed under 3-point bending conditions, resulting in the formation of shear bands before failure. The immediate environment of shear bands was investigated in detail using fluctuation electron microscopy to extract information on the strain-induced modifications of the medium-range order (MRO) and its lateral extension. Characteristic material-independent gradients were observed for the tensile and compressive sides of the samples indicating the impact of the local stress state on the MRO. Our results reveal an upper limit of a few microns for the lateral extension of the shear-affected environments of shear bands.
\end{abstract}


\end{frontmatter}


\section{Introduction}
Metallic glasses (MGs) are principally of great interest as structural materials since they exhibit high strength and hardness. However, most MGs lack ductility, especially under tension where zero ductility prevails. However, in recent years progress has been made in developing bulk metallic glasses (BMGs) exhibiting respectable ductility during cold rolling, bending and compression tests. Upon inhomogeneous deformation, that is at low temperatures and high stresses, the plasticity is manifested by a macroscopic sliding along a localized region called a shear band (SB) having a thickness of about 15 nm or less. It has been found that such SBs contain alternating density changes accompanied by structural changes in the medium range order (MRO) \cite{rosner2014density,schmidt2015quantitative,hieronymus2017shear,Hilke2019275}. The current understanding of how mesoscopic SBs evolve from shear transformation zones (STZs), which are regarded as the main carriers for plasticity in metallic glasses \cite{hassani2019probing}, is that alignments of Eshelby-like quadrupolar stress-field perturbations lead to percolation and thus to SB formation \cite{hieronymus2017shear,dasgupta2012microscopic}. The formation of a SB upon deformation creates an interface between the SB and the matrix. To maintain cohesion at the interface atoms need to be rearranged in the matrix, which consequently should also affect the adjacent matrix regions. Indeed, recent publications report on the existence of so-called shear band affected zones (SBAZs) \cite{cao2009structural,maass2014single,shen2018shear,kuchemann2018shear,liu2018elastic}. To address the relation between the structure of amorphous materials in terms of MRO and their mechanical behavior in more detail, the immediate environment of SBs in BMGs was investigated using fluctuation electron microscopy (FEM) to extract the local information on MRO \cite{treacy1996variable,voyles2002fluctuation,treacy2005fluctuation}. FEM contains information about the four-body correlation of atom pairs (pair-pair correlation function) $g_4(|\vec{r}_1|, |\vec{r}_2|, |\vec{r}|, \theta)$ yielding information about the MRO (cluster size and volume fraction) in amorphous materials \cite{treacy2005fluctuation,Bogle2010}. The obtained MRO profiles measured across SBs from tensile and compressive sides of 3-point bending tests display the impact of the local stress state on the MRO and thus shed more light on the lateral extension of deformation in SB environments.

\section{Methods}
FEM and high-angle annular dark-field scanning transmission electron microscopy (HAADF-STEM) were performed with a Thermo Fisher Scientific FEI Themis 300 G3 transmission electron microscope (TEM) operated at 300\,kV. Nanobeam diffraction patterns (NBDPs) were acquired with parallel illumination using a probe size of 1.3\,nm at full width half maximum (FWHM) in \textmu Probe-STEM mode operated at spotsize 8 with a 50\,\textmu m C2 aperture giving a semi-convergence angle of 0.8\,mrad. The probe size used for the acquisition of the NBDPs was measured directly on the Ceta camera prior to the FEM experiments using Digital Micrograph plugins by D. Mitchell \cite{mitchell2005scripting}. A beam current of 15\,pA and a camera length of 77\,mm were used for the acquisition in combination with an US 2000 CCD camera (Gatan) at binning 4 ($512\times512$ pixels). Data sets consisting of $60\times200$ (Fig.~\ref{fig:FIG2}) or $80\times200$ (Fig.~\ref{fig:FIG3}) individual NBDPs were performed across SBs from tensile and compressive sides of 3-point bending tests. From either 60 or 80 individual NBDPs containing the spatially resolved diffracted intensity $I(|\vec{k}|,\vec{r})$ at a fixed distance $\vec{r}$ from the SB, the normalized variance $V(|\vec{k}|,\vec{r})$ was calculated in the form of the annular mean of variance image ($\Omega_{VImage} (k)$) \cite{daulton2010nanobeam} using
\begin{equation}
V(|\vec{k}|,R = 1.3\,nm) = \frac{\left\langle I^2 (\vec{k},R,\vec{r}) \right\rangle}{\left\langle I (\vec{k},R,\vec{r}) \right\rangle^2}-1
\label{EQ:NVAR}
\end{equation}
where $\left\langle\,\,\right\rangle$ indicates the averaging over different sample positions r or volumes and R denotes the FWHM of the probe and thus the reciprocal space resolution. Since the normalized $V(k,R)$ scales with $1/t$ \cite{yi2011effect,radic2019fluctuation}, a thickness correction was made using the slope of the HAADF signal across the SB \cite{voyles2002fluctuatione}. The individual normalized variances were subsequently plotted against their position. The peak height of the first normalized variance peak was taken as a measure of the MRO volume fraction.

\section{Results}

\begin{figure}
	\includegraphics[width=\columnwidth]{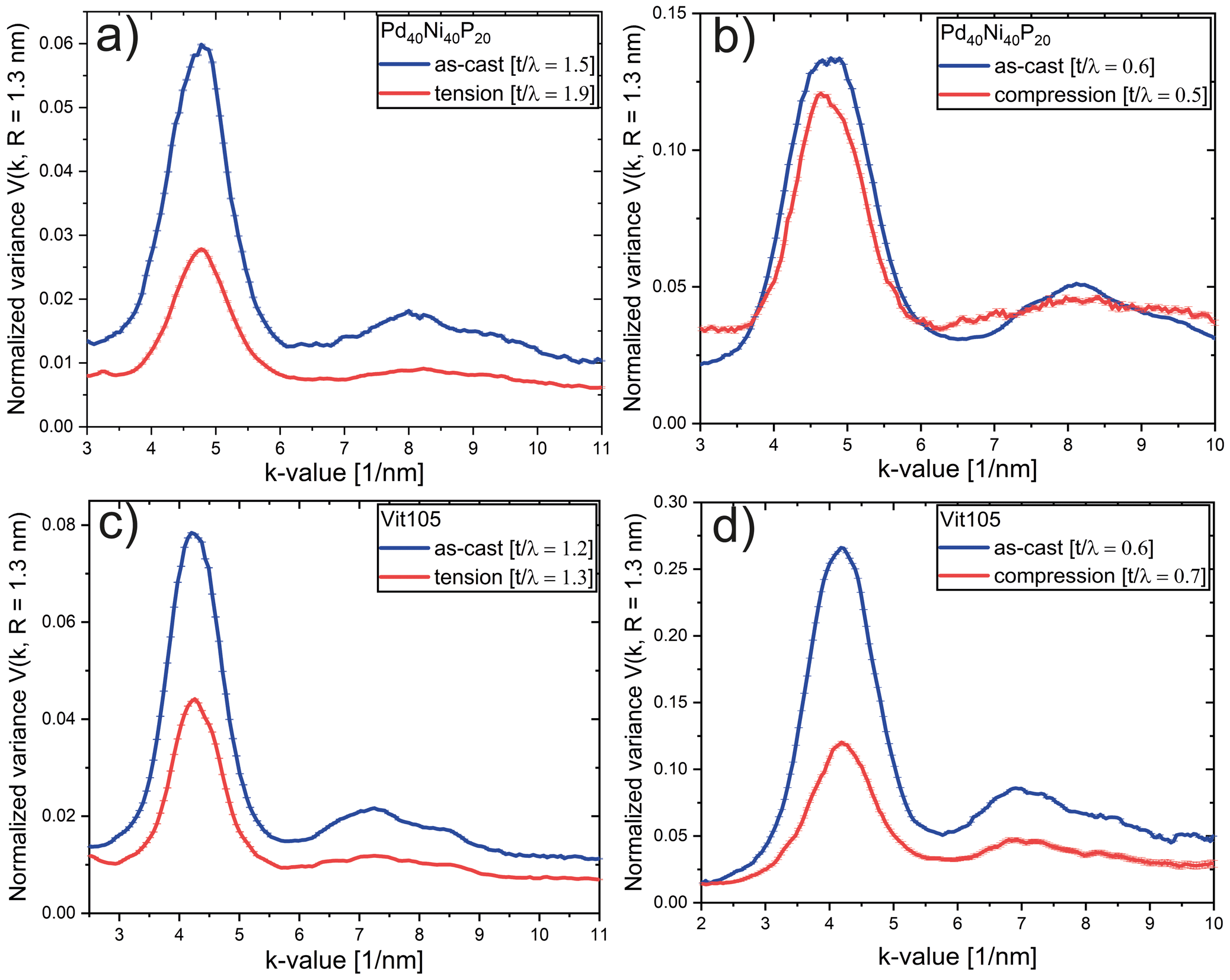}%
	\caption{Normalized variance profiles of as-cast and deformed states (compressive and tensile side) of Pd$_{40}$Ni$_{40}$P$_{20}$ (a, b) and Vit105 (c, d) showing a reduction in peak height due to deformation.}
	\label{fig:FIG1}
\end{figure}

3-point bending test of notched bars were carried out (see supplementary video in Appendix A). A more detailed description is given in reference \cite{geissler2019}. During such deformation tests, the area around the notch is dominated by tensile strain whereas the side opposite to the notch is mainly under compressive strain. In this paper these specific regions are referred to as the tensile and compressive sides, respectively. FIB lamellae containing SBs from each side (tensile and compressive) of the deformed samples were prepared perpendicular to shear steps penetrating through the surfaces. Fig.~\ref{fig:FIG1} shows representative examples of normalized variance profiles from as-cast and deformed matrix states (compressive and tensile side) of Pd$_{40}$Ni$_{40}$P$_{20}$ (Fig.~\ref{fig:FIG1}a,b) and Zr$_{52.5}$Cu$_{17.9}$Ni$_{14.6}$Al$_{10}$Ti$_{5}$ (Vit105) (Fig.~\ref{fig:FIG1}c,d). It is worth noting here that the normalized variance $V(k,R)$ scales with $1/t$, where $t$ is the foil thickness \cite{yi2011effect,radic2019fluctuation,voyles2002fluctuatione}. However, $V(k,R)$ signals of different materials acquired at similar foil thicknesses are directly comparable on an absolute scale. Comparing the peak heights of samples with similar foil thickness $t/\lambda = 0.6$ (Fig.~\ref{fig:FIG1}b,d) as a measure for the MRO volume fraction shows that the MRO in the as-cast state is higher for Vit105 than for Pd$_{40}$Ni$_{40}$P$_{20}$. Moreover, three of the individual variance profiles in Fig.~\ref{fig:FIG1} show a clear reduction in the peak height (by a factor of about 2); however it should be noted that the reduction is less pronounced for the compression side of Pd$_{40}$Ni$_{40}$P$_{20}$ as shown in Fig.~\ref{fig:FIG1}b. This behavior was also observed for severely deformed Pd$_{40}$Ni$_{40}$P$_{20}$ \cite{Zhou2019a,Shahkhali2019}.

\begin{figure}
	\centering
	\includegraphics[width=\columnwidth]{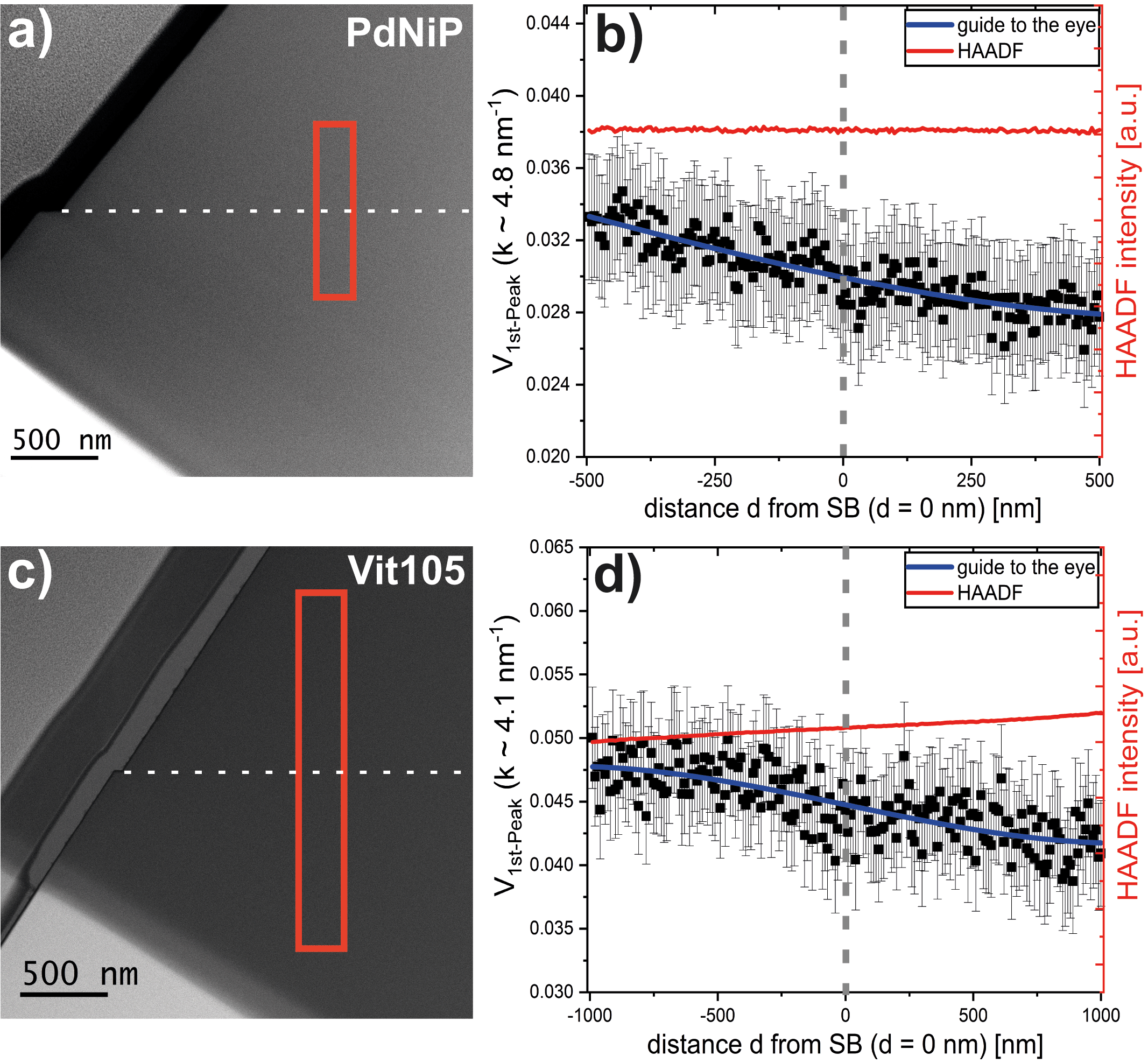}%
	\caption{HAADF-STEM image showing FIB-prepared lamellae with a shear band in (a) Pd$_{40}$Ni$_{40}$P$_{20}$ and (c) in Vit105 taken from the tensile side of the notched 3-point bending test. Note the shear steps at the surface. The red boxes indicate the regions from which the  individual NBDPs were acquired (from top to bottom). The NBDP map has a size of 200 nm x 1000 nm for Pd$_{40}$Ni$_{40}$P$_{20}$ (a) and 300 nm x 2000 nm for Vit105 (b) containing each 60 x 200 individual NBDPs. (b) and (d) display the height of the first normalized variance peaks, calculated from an avergage of 60 NBDPs, versus the distance from the shear band. Note the shallow gradient from left to right. The red profiles display the HAADF intensity indicating the foil thickness along the scanned area.}
	\label{fig:FIG2}
\end{figure}

The immediate environments of SBs were inspected by acquiring NBDPs over larger areas. Fig.~\ref{fig:FIG2}a and Fig.~\ref{fig:FIG2}c show HAADF-STEM images of the FIB-prepared lamellae. Since the contrast from the SBs is very faint, their positions are indicated by white dashed lines, ending at the surface steps. The FEM analyses were performed for SBs of the tensile sides in Pd$_{40}$Ni$_{40}$P$_{20}$ and Vit105 corresponding to the red rectangular areas shown in Fig.~\ref{fig:FIG2}a,c. The results; that is, the normalized variances mapped parallel to the dashed grey line are shown in Fig.~\ref{fig:FIG2}b and Fig.~\ref{fig:FIG2}d. A shallow continuous gradient is observed for both BMGs running from one side to the other (Fig.~\ref{fig:FIG2}b and Fig.~\ref{fig:FIG2}d). The peak height, which is taken as a measure for the MRO volume fraction, was reduced by a factor of about 2 after deformation (cf. Fig.~\ref{fig:FIG1}). It is worth noting that thickness effects (see red HAADF profile in Fig.~\ref{fig:FIG2}b and Fig.~\ref{fig:FIG2}d), potentially influencing these gradients in the variance peak heights, were eliminated by a thickness correction \cite{treacy2012examination}. In the case of Fig.~\ref{fig:FIG2}b the HAADF intensity is almost flat showing that there is hardly any influence from the foil thickness. Since there is no plateau level in the variance peak height visible, it is difficult to estimate the lateral extension of such shallow gradients. Fig.~\ref{fig:FIG2}d suggests the extension to be in the range of several microns.

\begin{figure}
	\centering
	\includegraphics[width=\columnwidth]{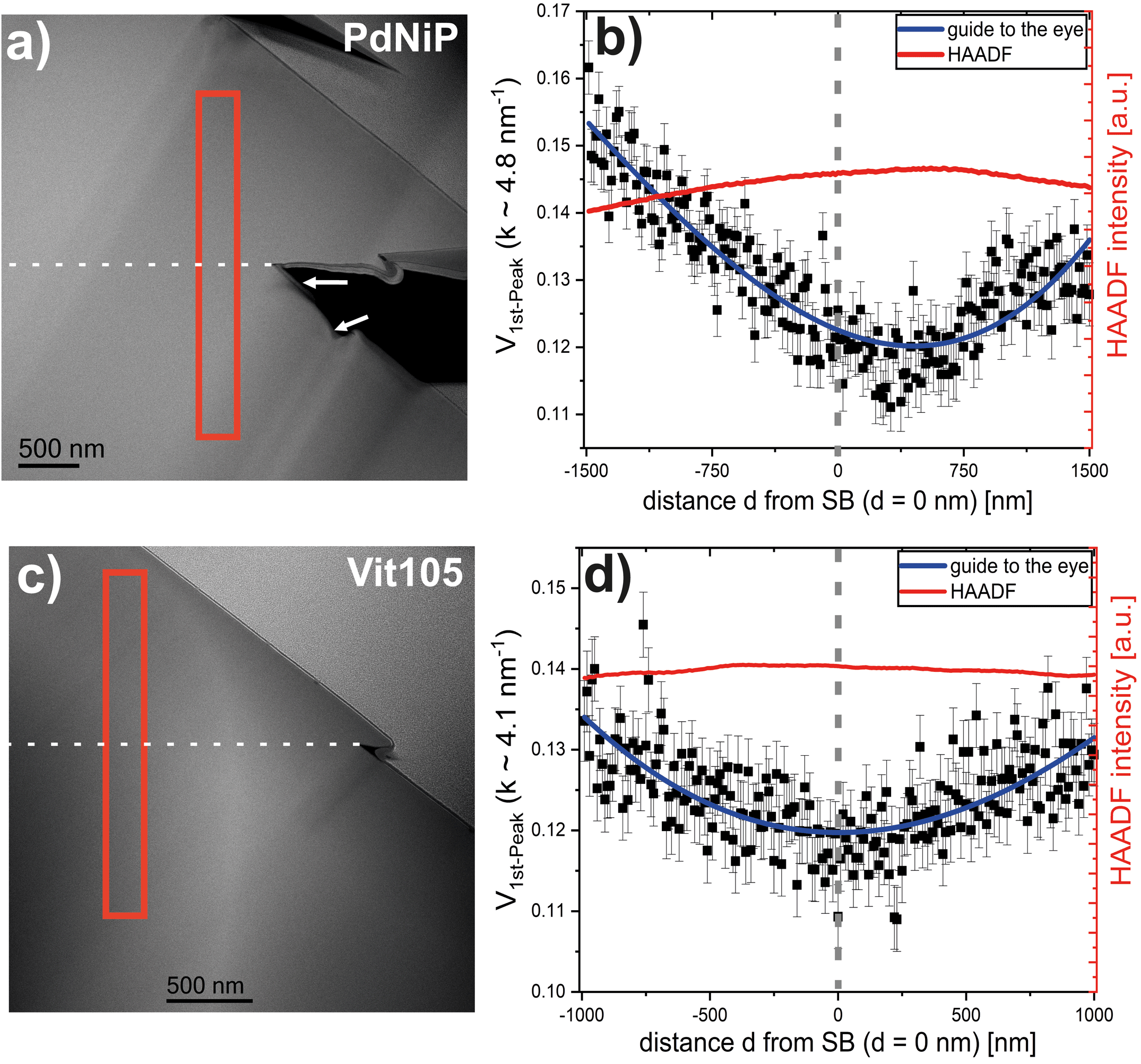}%
	\caption{HAADF-STEM image showing a shear band in (a) Pd$_{40}$Ni$_{40}$P$_{20}$ and (b) Vit 105 taken from the compressive side of the notched bars which were deformed by 3-point bending tests. Note the shear step at the surface in each case. Note the additional surface shear steps (see white arrows in (a)) indicating the existence of more SBs. The red boxes indicate the regions from which the individual NBDPs were acquired (from top to bottom). The NBDP map has a size of 300\,nm x 3000\,nm for Pd$_{40}$Ni$_{40}$P$_{20}$ (a) and 300\,nm x 2000\,nm for Vit105 (b) containing each $80\times200$ individual NBDPs. (b) and (d) display the height of the first normalized variance peaks, calculated from an avergage of 80 NBDPs, versus the distance from the shear band. The red profiles correspond to the HAADF intensity and provide a reference for the foil thickness along the scanned area.}
	\label{fig:FIG3}
\end{figure}

In like manner, the regions around the SBs from the compressive side were investigated. Fig.~\ref{fig:FIG3}a and Fig.~\ref{fig:FIG3}c depict HAADF-STEM images of SBs from the compressive side. The results of the FEM analyses are shown in Fig.~\ref{fig:FIG3}b and Fig.~\ref{fig:FIG3}d. In contrast to the tensile side, the measurements of the compressive side display for both BMGs a pronounced dip in the peak height of the normalized variance $V(k,R)$ in the vicinity of the SB. While a symmetric dip with a clear minimum at the SB position is observed for Vit105, the dip minimum for Pd$_{40}$Ni$_{40}$P$_{20}$ is displaced with respect to the SB position. The dip minimum identifies the highest stress concentration in the shear band leading to the greatest reduction of MRO. The lateral extension of the MRO curve is larger than 1\,\textmu m. The influence of the foil thickness is shown by the HAADF signal (red curves in Fig.~\ref{fig:FIG3}b and Fig.~\ref{fig:FIG3}d). Since a thickness correction was carried out on the variance peak heights, the dips in the MRO curves are not thickness artifacts. 

\section{Discussion}
In the following we discuss the robustness of this analysis.

\subsection{Error assessment of FEM analysis}
FEM analyses reveal differences in the absolute peak height of the normalized variance $V(k,R)$ in the vicinity of SBs \cite{Hilke2019275} with respect to the undeformed ''as-cast`` matrix by a factor of up to about 2 (Fig.~\ref{fig:FIG1}). Thus, by considering the absolute peak height of the normalized variance as a measure for the MRO volume fraction, changes in the MRO of the SB environment are detectable. Thickness gradients along the SB (parallel to the scan direction) are also an error source since the calculation of the normalized variances results from the averaging of the individual NBDPs along this direction. An error estimation due to such thickness effects (parallel to the SB) can be obtained from the match of very small or high k-values (see 3D plots in the Supplementary Material (Fig. S2)). These errors can be estimated to a maximum of $V = 0.005$ and are thus far below the observed absolute differences in peak height of the normalized variance $V(k,R)$. As the thickness effects perpendicular to the scan direction (across the SB) were already corrected using the HAADF intensities shown in Fig.~\ref{fig:FIG2} and Fig.~\ref{fig:FIG3}, the obtained MRO profiles are free from thickness artifacts.

\subsection{Variance peak height: variations and shapes}
The detected peak height variations of $V(k, R)$ reveal significant differences between the tensile and compressive side of the sample; that is, a pronounced dip of $V(k, R)$ at the SBs of the compressive side. For the tensile side a continuous shallow gradient seems to be characteristic. No clear minimum is visible. While a pronounced $V(k)$-dependence with a clear minimum at the SB position is observed for the compressive side of Vit105, the shape of the $V(k)$-relation for the compressive side of the SBs in Pd$_{40}$Ni$_{40}$P$_{20}$ is asymmetric, showing no overlap between the location of the SB and the minimum of the data. Intuitively, one would expect the minimum to overlap with the position of the SB since the MRO reduction in the SB should be highest. Hidden SBs explain the shift of the dip minimum since two more surface shear steps (see white arrows in Fig.~\ref{fig:FIG3}a) clearly indicate the presence of more SBs within the inspected SB area. 

\noindent Different structural changes due to the tension-compression asymmetry in BMGs have also been measured using high-energy synchrotron x-ray scattering \cite{chen2013atomic}. An explanation for the difference in shape of the $V(k)$ curves between compressive and tensile sides may be found in the anharmonicity of the interatomic potential so that atoms or clusters are differently affected at a given stress, i.e. pulling them apart is easier than pushing them together. This fits to the observation that the SBs formed earlier and more abundantly on the tensile side (see Supplementary Video).

Another point that needs to be discussed in connection with the shape of the $V(k)$ curves is the amount of deformation (material flow) carried by each individual SB. This can be estimated by the heights of the shear steps at the surface after deformation. Tab.~\ref{tab:TAB1} lists the heights of the shear steps of each SB investigated in this study. By comparing these data with the FEM results shown in Fig.~\ref{fig:FIG2} and Fig.~\ref{fig:FIG3}, there seems to be no correlation between the shear offset heights and the impact on the SB environment. 

Next, we discuss the influence of the testing geometry on the observed deformation behavior. The different geometrical constraint for tension and compression loading defines the complex stress states in the deformation experiments of BMGs (here notched 3-point bending test) and has thus great influence on the deformation behavior (ductile-brittle) \cite{wu2008strength,conner2003shear}. Since similar deformation signatures (MRO gradients in Fig.~\ref{fig:FIG2} and Fig.~\ref{fig:FIG3}) were found for both tested BMGs showing either ductile or brittle behavior, we believe that the observed MRO gradients represent characteristic stress states for the compressive and tensile side of notched 3-point bending tests. In fact, finite element simulations of BMGs under tensile load \cite{chen2013effect,chen2013deformation,chen2014deformation,chen2019effect} showed Von Mieses stress distributions which correspond to the shapes of the $V(K)$ curves observed for the tensile side (Fig.~\ref{fig:FIG2}). This means in conclusion that the MRO memorizes the impact of the stress field. 

\begin{table}[htbp]
	\caption{Shear step height originating from the investigated SBs.}
	\vspace{5pt}
		\begin{tabular}{c|c}
			Sample / location & Shear offset  \\
			 & height at surface \\
			\hline \hline
			Pd$_{40}$Ni$_{40}$P$_{20}$ [tensile side] & 130\,nm\\ 
			Vit105 [tensile side] & 40\,nm \\ 
			\hline
			Pd$_{40}$Ni$_{40}$P$_{20}$ [compressive side] & 1.7\,\textmu m\\ 
			Vit105 [compressive side] & 200\,nm\\ 
		\end{tabular} 
	\label{tab:TAB1}
\end{table}

\subsection{Shear-affected zones}
Our results, including previous strain analyses \cite{binkowski2015sub} on individual SBs, show that the SB affected zones are in the range of a few microns as an upper limit. However, shear band environments probed by nanoindentation (hardness), Young’s modulus measurements or changes in magnetic domains were reported to extend to $10-160$ microns \cite{cao2009structural,maass2014single,shen2018shear,kuchemann2018shear,liu2018elastic}. Moreover, these experiments also showed an increase in the lateral extension of the affected zones with in-creasing deformation \cite{maass2014single,shen2018shear,pan2011softening}. We see two reasons for this discrepancy: Firstly, the fact that the measurement scales (lateral resolution) are not comparable to each other. Secondly, deformation frequently leads to (hidden) multiple shear banding with SBs branching off the primary one \cite{csopu2020atomic}. Such shear band branches do not always penetrate through the surfaces and thus often remain undetected by more indirect methods (nanoindentation, atomic or magnetic force microscopy). In conclusion our results, having a very good lateral resolution with the ability to visualize the individual SBs, strongly suggest a lateral extension of shear-affected zones in the range of a few microns as an upper limit.

\section{Conclusions}
\noindent Fluctuation electron microscopy revealed a detailed structural picture of the interplay between deformation and MRO structure obtained from two representative BMGs (Pd$_{40}$Ni$_{40}$P$_{20}$ and Zr$_{52.5}$Cu$_{17.9}$Ni$_{14.6}$Al$_{10}$Ti$_{5}$) performed under 3-point bending conditions. (i) Prior to deformation, the amount of MRO was observed to be higher for Vit105 than for Pd$_{40}$Ni$_{40}$P$_{20}$. The degree of MRO was reduced after deformation. Profiling the MRO of shear band environments from compressive and tensile sides revealed characteristic gradients which seem to be material-independent and thus displaying the impact of the local stress state on the MRO. Our results show a lateral extension of shear-affected environments of shear bands in bulk metallic glasses with an upper limit of a few microns. 

\section{Acknowledgments}
We gratefully acknowledge financial support by the DFG via SPP 1594 (Topological engineering of ultra-strong glasses, WI 1899/27-2 and GE 1106/11) and WI 1899/29-1 (Coupling of irreversible plastic rearrangements and heterogeneity of the local structure during deformation of metallic glasses, projekt number 325408982). Moreover, DFG funding is acknowledged for TEM equipment via the Major Research Instrumentation Programme under INST 211/719-1 FUGG.


\bibliography{mybibfile}

\end{document}